\documentclass[twoside,twocolumn,9pt]{article}
\usepackage{extsizes}
\usepackage[super,sort&compress,comma]{natbib} 
\usepackage[version=3]{mhchem}
\usepackage[left=1.5cm, right=1.5cm, top=1.785cm, bottom=2.0cm]{geometry}
\usepackage{balance}
\usepackage{times,mathptmx}
\usepackage{sectsty}
\usepackage{graphicx} 
\usepackage{lastpage}
\usepackage[format=plain,justification=justified,singlelinecheck=false,font={stretch=1.125,small,sf},labelfont=bf,labelsep=space]{caption}
\usepackage{float}
\usepackage{fancyhdr}
\usepackage{fnpos}
\usepackage[english]{babel}
\addto{\captionsenglish}{%
  
}
\usepackage{array}
\usepackage{droidsans}
\usepackage{charter}
\usepackage[T1]{fontenc}
\usepackage[usenames,dvipsnames]{xcolor}
\usepackage{setspace}
\usepackage[compact]{titlesec}
\usepackage{hyperref}
\usepackage{epstopdf}%This line makes .eps figures into .pdf - please comment out if not required.

\definecolor{cream}{RGB}{222,217,201}

\begin{document}

\pagestyle{fancy}
\thispagestyle{plain}
\fancypagestyle{plain}{

%%%HEADER%%%
\fancyhead[C]{\includegraphics[width=18.5cm]{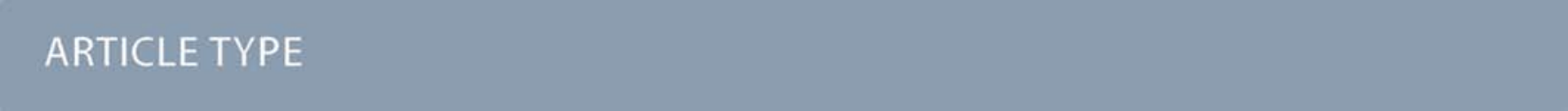}}
\fancyhead[L]{\hspace{0cm}\vspace{1.5cm}\includegraphics[height=30pt]{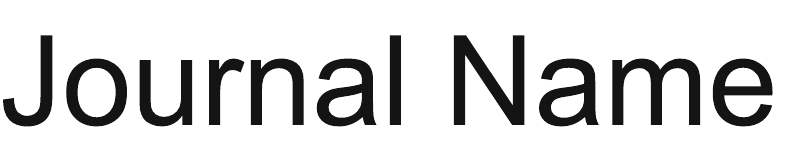}}
\fancyhead[R]{\hspace{0cm}\vspace{1.7cm}\includegraphics[height=55pt]{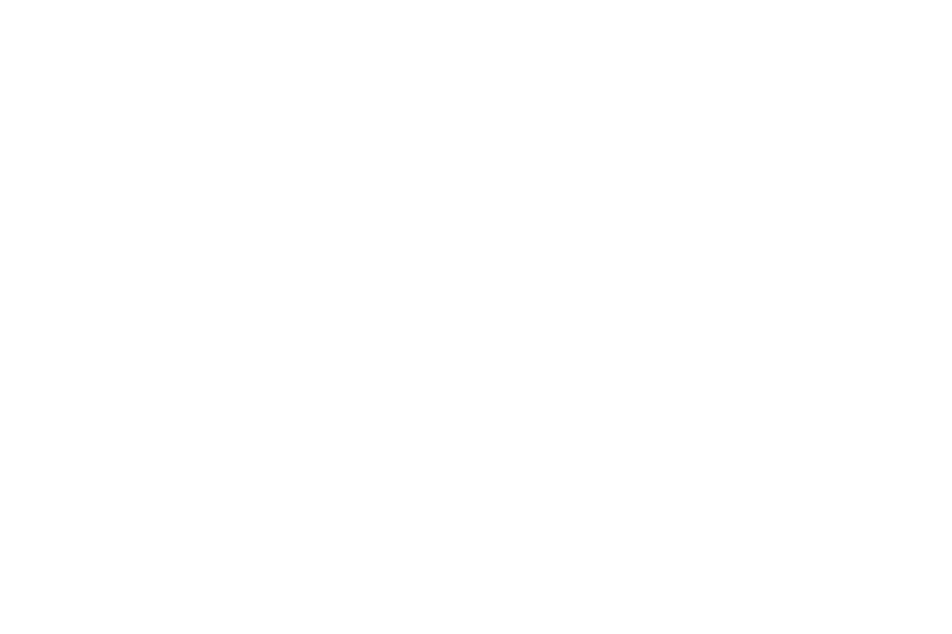}}
\renewcommand{\headrulewidth}{0pt}
}
%%%END OF HEADER%%%

%%%PAGE SETUP - Please do not change any commands within this section%%%
\makeFNbottom
\makeatletter
\renewcommand\LARGE{\@setfontsize\LARGE{15pt}{17}}
\renewcommand\Large{\@setfontsize\Large{12pt}{14}}
\renewcommand\large{\@setfontsize\large{10pt}{12}}
\renewcommand\footnotesize{\@setfontsize\footnotesize{7pt}{10}}
\makeatother

\renewcommand{\thefootnote}{\fnsymbol{footnote}}
\renewcommand\footnoterule{\vspace*{1pt}% 
\color{cream}\hrule width 3.5in height 0.4pt \color{black}\vspace*{5pt}} 
\setcounter{secnumdepth}{5}

\makeatletter 
\renewcommand\@biblabel[1]{#1}            
\renewcommand\@makefntext[1]% 
{\noindent\makebox[0pt][r]{\@thefnmark\,}#1}
\makeatother 
\renewcommand{\figurename}{\small{Fig.}~}
\sectionfont{\sffamily\Large}
\subsectionfont{\normalsize}
\subsubsectionfont{\bf}
\setstretch{1.125} %In particular, please do not alter this line.
\setlength{\skip\footins}{0.8cm}
\setlength{\footnotesep}{0.25cm}
\setlength{\jot}{10pt}
\titlespacing*{\section}{0pt}{4pt}{4pt}
\titlespacing*{\subsection}{0pt}{15pt}{1pt}
%%%END OF PAGE SETUP%%%

%%%FOOTER%%%
\fancyfoot{}
\fancyfoot[LO,RE]{\vspace{-7.1pt}\includegraphics[height=9pt]{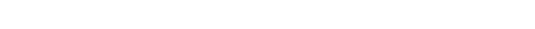}}
\fancyfoot[CO]{\vspace{-7.1pt}\hspace{13.2cm}\includegraphics{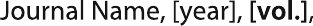}}
\fancyfoot[CE]{\vspace{-7.2pt}\hspace{-14.2cm}\includegraphics{RF}}
\fancyfoot[RO]{\footnotesize{\sffamily{1--\pageref{LastPage} ~\textbar  \hspace{2pt}\thepage}}}
\fancyfoot[LE]{\footnotesize{\sffamily{\thepage~\textbar\hspace{3.45cm} 1--\pageref{LastPage}}}}
\fancyhead{}
\renewcommand{\headrulewidth}{0pt} 
\renewcommand{\footrulewidth}{0pt}
\setlength{\arrayrulewidth}{1pt}
\setlength{\columnsep}{6.5mm}
\setlength\bibsep{1pt}
%%%END OF FOOTER%%%

%%%FIGURE SETUP - please do not change any commands within this section%%%
\makeatletter 
\newlength{\figrulesep} 
\setlength{\figrulesep}{0.5\textfloatsep} 

\newcommand{\topfigrule}{\vspace*{-1pt}% 
\noindent{\color{cream}\rule[-\figrulesep]{\columnwidth}{1.5pt}} }

\newcommand{\botfigrule}{\vspace*{-2pt}% 
\noindent{\color{cream}\rule[\figrulesep]{\columnwidth}{1.5pt}} }

\newcommand{\dblfigrule}{\vspace*{-1pt}% 
\noindent{\color{cream}\rule[-\figrulesep]{\textwidth}{1.5pt}} }

\makeatother
%%%END OF FIGURE SETUP%%%

%%%TITLE, AUTHORS AND ABSTRACT%%%
\twocolumn[
  \begin{@twocolumnfalse}
\vspace{3cm}
\sffamily
\begin{tabular}{m{4.5cm} p{13.5cm} }

\includegraphics{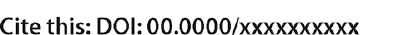} & \noindent\LARGE{\textbf{Inverse Leidenfrost drop manipulation using menisci}} \\%Article title goes here instead of the text "This is the title"
\vspace{0.3cm} & \vspace{0.3cm} \\

 & \noindent\large{Ana\"\i s Gauthier $^{\ast}$\textit{$^{a}$}, Guillaume Lajoinie\textit{$^{a}$}, Jacco H. Snoeijer\textit{$^{a}$} and Devaraj van der Meer\textit{$^{a}$}} \\

\includegraphics{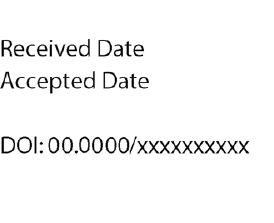} & \noindent\normalsize{Drops deposited on an evaporating liquid bath can be maintained in an inverse Leidenfrost state by the vapor emanating from the bath, making them levitate and hover almost without friction. These perfectly non-wetting droplets create a depression in the liquid interface that sustains their weight, which generates repellent forces when they approach a meniscus rising against a wall. Here, we study this reflection in detail, and show that frictionless Leidenfrost drops are a simple and efficient tool to probe the shape of an unknown interface. We then use the menisci to control the motion of the otherwise elusive drops. We create waveguides to direct and accelerate them and use parabolic walls to reflect and focus them. This could be particularly beneficial in the scale up of droplet cryopreservation processes: capillary interactions can be used to transport, gather and collect vitrified biological samples in absence of contact and contamination.
} \\

\end{tabular}

 \end{@twocolumnfalse} \vspace{0.6cm}

  ]
%%%END OF TITLE, AUTHORS AND ABSTRACT%%%

%%%FONT SETUP - please do not change any commands within this section
\renewcommand*\rmdefault{bch}\normalfont\upshape
\rmfamily
\section*{}
\vspace{-1cm}

%%%FOOTNOTES%%%

\footnotetext{\textit{$^{a}$~Physics of Fluids group and Max Plank Center Twente. Mesa + Institute and Faculty of Science and Technology, J.M. Burgers Centre for Fluid Dynamics and Max Plank Center Twente for Complex Fluid Dynamics. University of Twente, P.O. Box 217 7500 AE Enschede, The Netherlands}}
%\footnotetext{\textit{$^{b}$~Address, Address, Town, Country. }}

%Please use \dag to cite the ESI in the main text of the article.
%If you article does not have ESI please remove the the \dag symbol from the title and the footnotetext below.
%\footnotetext{\dag~Electronic Supplementary Information (ESI) available: [details of any supplementary information available should be included here]. See DOI: 00.0000/00000000.}
%%additional addresses can be cited as above using the lower-case letters, c, d, e... If all authors are from the same address, no letter is required
%
%\footnotetext{\ddag~Additional footnotes to the title and authors can be included \textit{e.g.}\ `Present address:' or `These authors contributed equally to this work' as above using the symbols: \ddag, \textsection, and \P. Please place the appropriate symbol next to the author's name and include a \texttt{\textbackslash footnotetext} entry in the the correct place in the list.}

%%%END OF FOOTNOTES%%%

%%%MAIN TEXT%%%%

%%%%%%%%%%%%%%%%%%
% \section{Introduction}
%%%%%%%%%%%%%%%%%%

Droplets deposited on a cryogenic liquid, such as liquid nitrogen can be levitated through the action of an insulating vapor layer originating from the bath \cite{Song:2010, Adda-Bedia:2016,Gauthier:2019}. This phenomenon is called the inverse Leidenfrost effect, by comparison with the usual Leidenfrost situation where volatile liquids are levitated above a hot solid \cite{Leidenfrost:1756, Biance:2003}. In both cases, the levitating particles are extremely mobile \cite{LeMerrer:2011, Hale:2016}, and controlling their motion is a challenge \cite{Maquet:2016}. Solid substrates can be sculpted to control drop motion, either macroscopically by giving them a curved shaped \cite{Ma:2017} (e.g. similarly to the spoon used by J.G. Leidenfrost \cite{Leidenfrost:1756}) or at a smaller scale by using textures to redirect the vapor below the drop \cite{Linke:2006,Cousins:2012}. Such strategies have not been proposed on liquids, where they are difficult to implement. Nonetheless, liquids can be reshaped up to the millimeter scale, using surface tension forces. In particular, the liquid interface distorts and creates a meniscus in the presence of a wall. For wetting liquids (with a contact angle $\theta<90^{\circ}$) the liquid rises along solid interfaces over a characteristic distance equal to the capillary length $a$. Floating objects interact with these menisci, often in an attractive way: small bubbles are seen to drift towards the edge of a glass \cite{Nicolson:1949} and particles (colloids, plant seeds) spontaneously climb them \cite{Peruzzo:2013} and even spontaneously self-assemble \cite{Cavallaro:2011,Ershov:2013}. On the contrary, hydrophobic objects such as liquid marbles \cite{Bormashenko:2015, Ooi:2015} or insects \cite{Vella:2005, Hu:2005, Liu:2007, Su:2010, Darmanin:2015} are repelled by wetting menisci. Levitating drops, which are perfectly non-wetting also bounce away from (wetting) menisci \cite{Hale:2019}, which has successfully been used to confine Leidenfrost droplets to a specific location \cite{Maquet:2016}. 

\noindent In this paper, we propose to use the repellency of inverse Leidenfrost droplets by menisci as a way to finely tune and control drop motion. To this end, we first model the droplet interaction with a single wall and subsequently combine multiple structures to enforce particle motion in a controlled direction, accelerate them and focus them onto a chosen location. Our systems might be useful in droplet cryopreservation processes, where droplets containing biological materials are vitrified on a liquid nitrogen bath. We demonstrate here that capillarity can be used to efficiently transport and collect multiple samples in absence of any contact and thus avoiding contamination.

\section{Experiment.} 
\begin{figure}[!h]% Use the figure* environment to get a wide figure \texttt{twocolumn} formatting. 
\centering
\includegraphics[width=0.99\columnwidth]{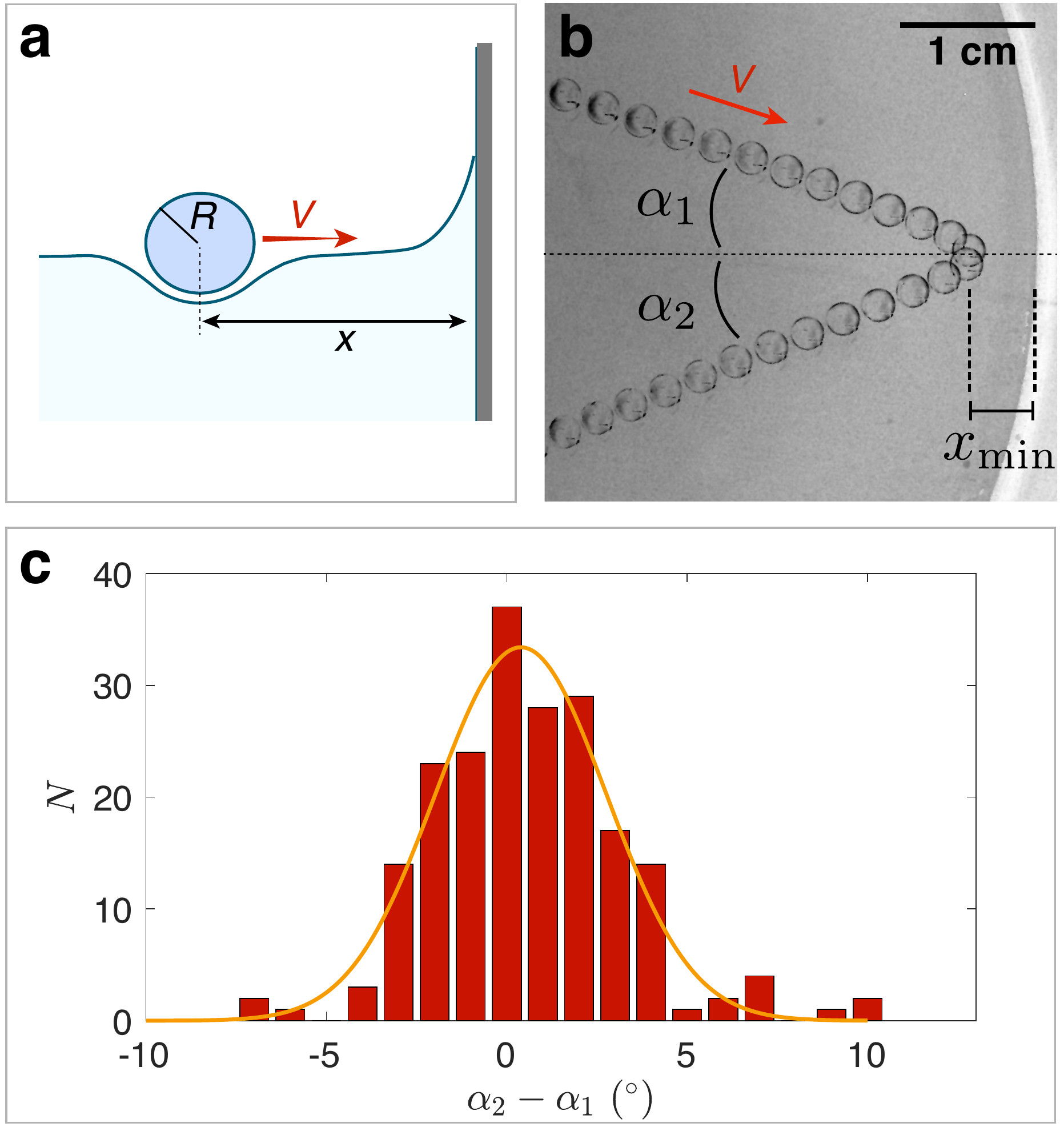}% Here is how to import EPS art
\caption{\label{figure1}\textbf{a.} A silicone oil drop, with radius $R$ is maintained in levitation above a liquid nitrogen bath. It hovers almost without friction: we consider its motion and dynamics as it approaches a wall. \textbf{b.} Top view of an experiment. A drop with radius $R = 1.2$ mm is repelled by the wetting meniscus against a wall. We name $\alpha_1$ the incoming angle, and $\alpha_2$ the reflected angle. See also Suppementary Movie 1. \textbf{c.} Difference $\alpha_2 - \alpha_1$ measured for more 200 rebounds: the drops are almost always perfectly reflected, with  $|\alpha_2 - \alpha_1| < 5^{\circ}$. The orange line is a Gaussian fit, with standard deviation $\sigma = 2.4^{\circ}$.}
\end{figure}

Millimeter-sized silicone oil droplets (with density $\rho$ = 930 kg/m$^3$ and radius $R$ between 0.8 and 2.0 mm) are released a few centimeters above a still liquid nitrogen bath, using calibrated needles. Liquid nitrogen is kept in a beaker with a diameter of 10 cm, and insulated (using a sacrificial bath \cite{Adda-Bedia:2016}) to keep it from boiling. Due to the large temperature gradient between the drops (initially at ambient temperature) and the bath (at its boiling temperature, -196$^{\circ}$C) nitrogen evaporates rapidly below the drops. The nitrogen flow maintains drops in levitation, in an ``inverse Leidenfrost" state \cite{Song:2010, Adda-Bedia:2016, Feng:2018, Gauthier:2019}. In absence of direct contact with the liquid surface that sustains them, the drops motion is only limited by friction forces arising from shear within the vapor film \cite{Biance:2003, Hale:2016, Gauthier:2019}. These forces are more than ten times smaller than the usual Stokes drag. In addition, a self-propulsion mechanism almost perfectly compensates the remaining friction, making these objects almost frictionless. This state arises from a spontaneous symmetry breaking within the film below the drop that partially redirects the nitrogen vapor flow, dragging the drop along with it \cite{Gauthier:2019}. The frictionless drops then move in straight lines (at a velocity $V$) along a direction randomly set by the initial symmetry breaking. The droplets deform the surface of the bath by their weight, generating a local depression of the interface, as illustrated in Figure \ref{figure1}a. We consider the interaction of an incoming droplet (with radius $R$ and velocity $V$) with a wall. Experiments are optically recorded from the top, using a high speed camera (Photron Mini UX-100) at a typical framerate of 500 fps, and the drops are tracked using an in-house Python algorithm. 

%%%%%%%%%%%%%%%%%% FIGURE 1 %%%%%%%%%%%%%%%%%%%%%%%%%%%%%%
%%%%%%%%%%%%%%%

\noindent
 Figure \ref{figure1}b and Supplementary Movie 1 present a top-view recording of a typical bouncing experiment, with an interframe time of 60 ms. A drop (with radius $R$ = 1.2 mm) approaches the glass wall of the beaker (in white) and is repelled from it. The particle has an initial velocity $V$ = 3.5 $\pm$ 0.1 cm/s (indicated with a red arrow) and approaches the wall with an initial angle $\alpha_1 = 20.0\,\pm\,0.5$ degrees measured with respect to the normal to the wall. From Figure \ref{figure1}b one can infer that the rebound is close to a perfect reflection, with a reflected angle $\alpha_2 = 21.5\,\pm\,0.5$ degrees. By measuring the difference between the incoming and reflected angles $\alpha_2 - \alpha_1$ for more than 200 bouncing events (with $\alpha_1$ varying between 0.5 and 30$^\circ$), we show in Figure \ref{figure1}c that this is a general feature of our system. The orange line is the best Gaussian fit to the experimental data, with a standard deviation $\sigma = 2.4^\circ$.
\noindent In addition, figure \ref{figure1}b shows that during the rebound, the drops do not come in contact with the wall but come to a stop at a distance $x = x_{\rm min}$ from it before bouncing back. In figure \ref{figure1}b, $x_{\rm min}$ = 3.5 $\pm$ 0.1 mm, a distance that is of the order of the diameter of the incoming object.
%The drop dynamics during the rebound are obtained by following the drop center of mass. An example is represented in Figure \ref{figure1}c, where we plot the velocity $V$ of a drop as a function of time $t$. The two different colors correspond to two drop with sizes $R$ = 1.4 mm and $R$ = 0.8 mm. The drops velocities $V$ before and after impact are constant, indicating that friction is indeed completely negligible here. In addition, the rebound itself (which lasts typically 200ms) is perfectly elastic, and the drops velocities after reflection are nearly identical to their velocity prior to it. As expected, the two plots overlap, which indicates that the bouncing dynamics are independent on the drop size. 

\smallskip

\section{Wall rebounds.}

\noindent \textbf{Capillary interaction.} We interpret these results as a consequence of the meniscus repellency. Liquid nitrogen, with surface tension $\gamma = 8.85\;10^{-3}$ N/m rises along the glass wall and generates a wetting meniscus (Figure \ref{figure1}a). Locally, the shape of the liquid surface $h(x)$ (with $x$ the distance to the wall) is determined by an equilibrium between the pressure change across the interface due to surface tension, and the hydrostatic pressure difference due to the rise of the liquid. This is expressed by the Young-Laplace equation:

\begin{equation}
\frac{d^2h}{dx^2}\left[1+\left(\frac{dh}{dx}\right)^2\right]^{-3/2} = \frac{\rho_{N} g h}{\gamma} = \frac{h}{a^2}
\label{Young-Laplace}
\end{equation}

\noindent where $g$ stands for gravity, $\rho_{N}$ for the density of liquid nitrogen and $a = \sqrt{\gamma/\rho_Ng}$ = 1.1 mm is the capillary length of liquid nitrogen. This equation has the analytic solution: \cite{Landau:1987, deGennes:2013}

\begin{equation}
x - x_0 = a\,\text{acosh}\left(\frac{2a}{h}\right) - 2a \left(1-\frac{h^2}{4a^2}\right)^{1/2}
\label{x}
\end{equation}

\noindent with $x_0$ an integration constant, chosen so that $h(x=x_0) = h^*$ where $h^*$ equals the height of the meniscus at the wall. For small deformations (in particular, far enough from the wall), equation \ref{Young-Laplace} can be simplified to $\frac{d^2h}{dx^2} = \frac{h}{a^2}$, which yields:
\begin{equation}
h(x) = a \cot\beta \exp^{-x/a}.
\label{h_simplifie}
\end{equation}
\noindent Here, $\beta$ is a constant that can be seen as an \textit{apparent} contact angle, obtained by matching the small slope approximation close to the wall. However, the small slope approximation does not necessarily hold close to the wall, especially for wetting liquids, and the value of $\beta$ is expected to overestimate the actual contact angle $\theta$, as we will see later.

\noindent When a particle approaches the wall, interaction between the depression of the interface around the drop and the wetting meniscus at the wall induces a capillary interaction potential $E_c$, which, following \cite{Vella:2005}, we estimate as the product of the effective weight of the particle $W$ with the meniscus shape $h(x)$. For perfectly non-wetting particles ($\theta = 180^{\circ}$), $W$ simplifies to $W = mg\frac{\rho}{\rho_N}$ with $m$ the drop mass, so that:

\begin{equation}
E_c(x) = mg\frac{\rho}{\rho_N} h(x).
\label{E_c}
\end{equation}
Note that $E_c(x)>0$ here, which indicates repellency.

\medskip

%%%%%%%%%% FIGURE 2 %%%%%%%%%%
\begin{figure*}% Use the figure* environment to get a wide figure \texttt{twocolumn} formatting. 
\centering
\includegraphics[width=1.9\columnwidth]{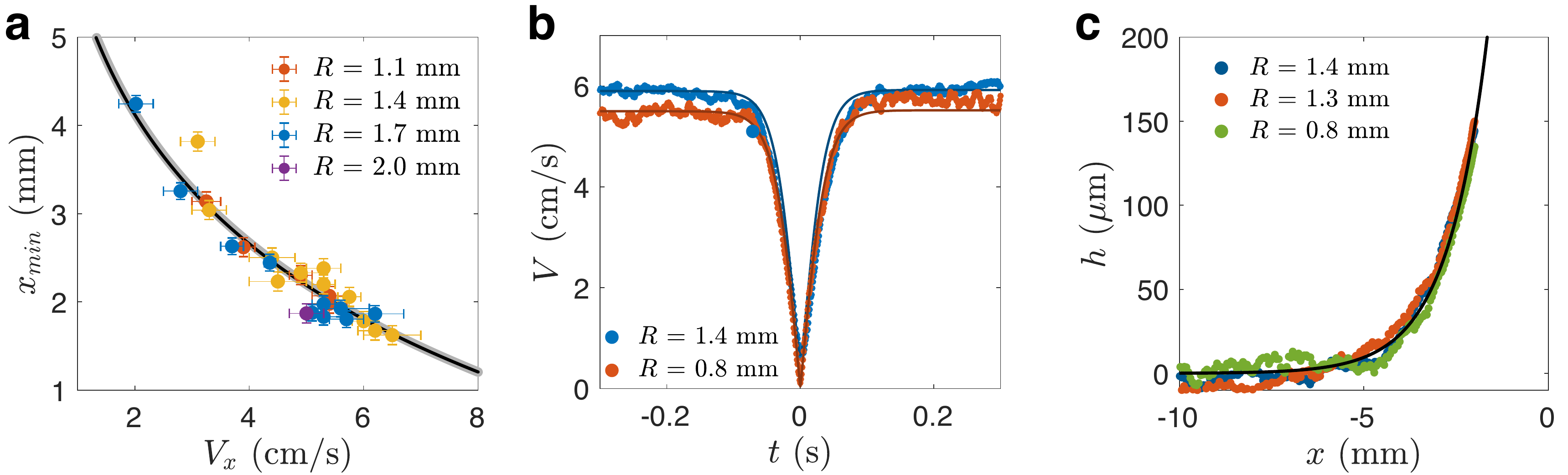}
\caption{\label{figure2}\textbf{a}. Minimal distance $x_{\rm min}$ between the drop center and the wall for multiple bouncing events, as a function of its incoming velocity $V_x$ and for varying drop sizes (1.1 mm $< R <$ 2 mm). The black line is the theoretical plot, from equation \ref{x} and constant $x_0$ = 0.45 mm, and the larger gray line is the small-slope approximation from equation \ref{h_simplifie}. \textbf{b.} Particle velocity $V$ as a function of time $t$ at it is repelled by the wall. The blue and red dots (respectively for $R = 1.4$ mm and $R$ = 0.8 mm) are the experiments, and the lines show the model. \textbf{c.} Comparison between the theoretical meniscus shape $h(x)$, in black and the shape reconstructed from the droplet dynamics (colored dots) for three different drop radii.}
\end{figure*}
%%%%%%%

\noindent \textbf{Bouncing dynamics.} In our experiment, the particles are almost perfectly frictionless, so that the kinetic energy of an approaching droplet $E_k = \frac{1}{2} m V^2$ is simply converted into potential energy as it climbs the liquid slide. The particle turns back when its velocity perpendicular to the wall is zero, at a position $x = x_{\rm min}$ (as seen in Figure \ref{figure1}b). Energy conservation between the positions $x = +\infty$ and $x = x_{\rm min}$ simply implies $\frac{1}{2}m V_{x}^2 = mg\frac{\rho}{\rho_N} h(x_{\rm min})$, with $V_{x}$ the initial velocity of the particule normal to the wall. Using equation \ref{x}, this gives the expression of $x_{\rm min}$:

\begin{eqnarray}
\label{x_min}
x_{min} &=& x_0 + a\,\text{acosh}\left(\frac{2a}{h_{\rm min}}\right) - 2a \left(1-\frac{h_{\rm min}^2}{4a^2}\right)^{1/2} \\ 
\text{with    }\;h_{\rm min} &=& \frac{V_{x}^2}{2g} \left(\frac{\rho}{\rho_N}\right)
\end{eqnarray}

\noindent In Figure \ref{figure2}a, we report the experimental measurements of $x_{\rm min}$ as a function of the drop velocity $V_x$ normal to the wall. As predicted by equation \ref{x_min}, $x_{\rm min}$ does not depend on the drop size, and the data obtained for varying $R$ (1.1 mm $<$ $R$ $<$ 2 mm) overlap. The solid black line shows equation \ref{x_min}, which matches the experimental data, with a fitting parameter $x_0$ = 0.45 mm. Interestingly, the small-slope approximation (Equation \ref{h_simplifie}, in gray) perfectly overlaps with the exact solution of the Young-Laplace equation (Equation \ref{x}, in black): in the region reached by the droplets, the small slope approximation holds. The fitting parameter $\beta = 50^\circ$ (the apparent contact angle) differs however from the actual contact angle $\theta = 30 \pm 5$ degrees, obtained by reporting the value of $x_0$ measured experimentally in eq. \ref{x}. Indeed, close to $x = 0$, equation \ref{h_simplifie} underestimates the meniscus deformation, and therefore overestimates the contact angle at the wall.
The dynamics of droplet rebounds give nonetheless a first evaluation of the contact angle $\theta$ of liquid nitrogen, using equation \ref{x}. Interestingly, constant evaporation of liquid nitrogen (a perfectly wetting liquid) close to the wall is associated with a contact angle $\theta$ as predicted by \cite{Morris:2001}, which value depends on the local heat and evaporation fluxes.

\smallskip

\noindent The capillary interaction model is further confirmed by following the center of mass of the particles as they bounce off the wall. In Figure \ref{figure2}b, we plot the velocity $V$ of two drops with sizes $R$ = 1.4 mm (blue dots) and $R$ = 0.8 mm (red dots) as a function of time $t$. The velocities $V$ before and after impact are identical and the rebound itself (which lasts typically 200 ms) is perfectly elastic, indicating that friction is indeed negligible here. Using equations \ref{h_simplifie} and \ref{E_c}, we integrated the equations of motion of the two droplets, and plotted the resulting velocities $V(t)$ as a slightly darker line on top of the experimental data of Figure \ref{figure2}b. The model nicely matches the experiments, indicating that the droplets dynamics can be fully modelled with a single repulsive capillary force $\vec{F} = -dE_c/dx\,\vec{e}_x$. $\vec{F}$ is normal to the wall, which also explains why the rebounds are close to perfect reflections, as seen in Figure \ref{figure1}c.

\smallskip

\noindent These results show that in absence of friction, non-wetting particles behave as capillary probes: one can infer from their dynamics any meniscus shape $h(x)$ through the capillary interaction potential. We demonstrate this in Figure \ref{figure2}c, where we compare the meniscus height calculated from the droplet dynamics $h(x) = \frac{1}{2g}\left(\frac{\rho_N}{\rho}\right) \left(V^2(x = \infty)-V^2(x)\right)$ to its theoretical value (in black, equation \ref{x}). The plots obtained for three different drop sizes are presented: they all overlap with the theoretical shape $h(x)$, demonstrating that our droplets indeed behave as interfacial probes.

%%%%% FIGURE 3 %%%%%%
\begin{figure*}[!h]% Use the figure* environment to get a wide figure \texttt{twocolumn} formatting. width=0.99\columnwidth
\centering
\includegraphics[width=1.9\columnwidth]{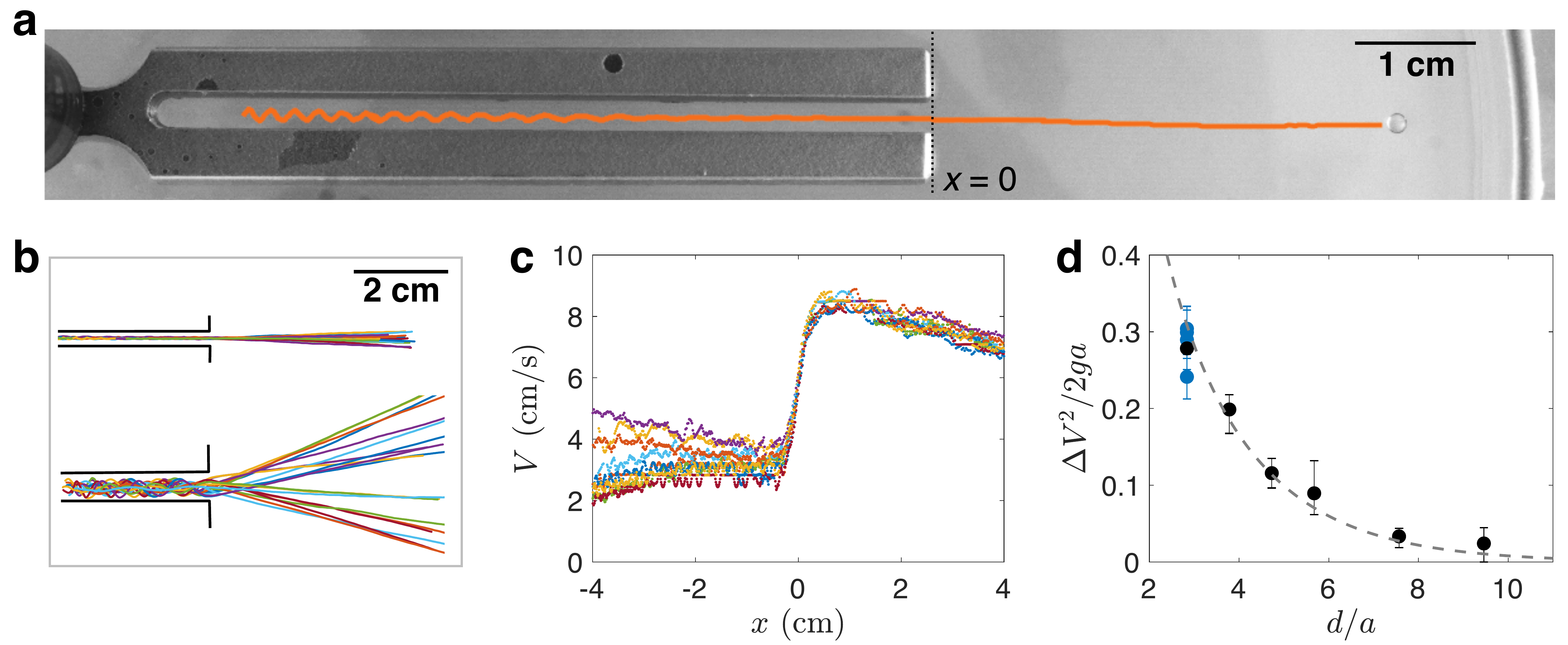}% Here is how to import EPS art
\caption{\label{figure3}\textbf{a.} Top view of a droplet gun experiment. A silicone oil drop with radius $R$ = 900 $\mu$m is deposited into an aluminium channel (with size $d$ = 3 mm) semi-immersed in the bath. Its trajectory is plotted in orange. After a few bounces against the walls, the drop is directed along the channel. See also Supplementary Movie 2. \textbf{b.} Trajectories of 20 drops (with radius $R$ = 900 $\mu$m) outside of two channels, with $d$ = 3 mm (top) and $d$ = 6 mm (bottom). \textbf{c.} Drop velocity $V$ inside ($x<0$) and outside ($x>0$) of a channel with width $d$ = 3 mm for 10 experiments. The drops slide along the meniscus and their velocity is more than doubled when they escape. \textbf{d.} Non-dimensional increase of the droplet kinetic energy, for varying channel size $d$ (black points) and drop radii (in blue). The dotted line is the theoretical model.}
\end{figure*}
\noindent
\section{Drop manipulation} In the following, we aim at directing the drop motion by using the menisci of well-defined wall geometries. We illustrate this with two examples: a droplet gun created by means of confinement within a channel and droplet focusing using parabolic walls.

\smallskip
\noindent \textbf{Channels.} In Figure 3a a drop (of radius $R$ = 0.9 mm) is deposited inside a narrow channel (with width $d$ = 3 mm). The orange line depicts the drop trajectory after its deposition: the drop, repelled by both walls, bounces back and forth a large number of times before the very small dissipation is experienced and stabilizes it at the center. The drop is thereby forced to slide in the direction of the channel. As visible in Figure \ref{figure3}a and Supplementary Movie 2, the particle maintains the enforced direction for tens of centimeters after escaping the channel. Figure \ref{figure3}b shows the superposition of twenty drop trajectories before and after escaping two channels with width $d$ = 3 mm and $d$ = 6 mm. When the channel size and the drop size are comparable (top image) the drops are almost perfectly guided: the spreading angle (measured at the exit of the channel) equals 1.6$^{\circ}$. The same drop deposited in a wider channel, however, will still be guided but less efficiently: for $d$ = 6 mm, the spreading angle reaches 30 degrees (figure \ref{figure3}b, bottom image).

\noindent Interestingly, the channels do not only guide the drops, but also accelerate them in a controlled and reproducible manner. Figure \ref{figure3}c shows the drop velocities $V$ inside ($x<0$) and outside ($x>0$) the channel for ten identical experiments. Inside the channel, the drop velocities $V_i$ are on the order of 3 cm/s, but $V$ is multiplied by almost a factor 3 when the drops escapes with $V_f \simeq 8.5$ cm/s. In our experiment, the velocity $V_f$ is higher than the terminal velocity $v_0 \simeq$ 5 cm/s) of the self-propelled droplets and friction causes a slow deceleration visible for $x > 1$ cm. In figure \ref{figure3}d, we report the non-dimensional variation of the kinetic energy of the drop $\Delta V^2/2ga$ with $\Delta V^2 = V_f^2-V_i^2$ for varying channel widths $d$ (3 mm $< d <$ 10 mm, black dots), and drop sizes (0.7 mm $< R <$ 1.3 mm, in blue). As the channel size increases, the drops are less and less accelerated until nothing happens for $d/a >$ 10. In addition, the data obtained for varying drop radii almost perfectly overlap in Figure \ref{figure3}d, indicating that the amplitude of the acceleration does not depend on the drop size.

\noindent What causes the acceleration? The distance $d$ between the walls is comparable to the capillary length, allowing the two menisci on each side to interact with each other, inducing a capillary rise in the channel. The height $H$ of liquid nitrogen inside of the channel is then simply determined by a 2D Jurin's law, and only depends on the distance $d$ between the two walls and on the contact angle at the wall. To simplify, we use the small-slope approximation of Young-Laplace equation, which, as demonstrated before, matches well the shape of the meniscus felt by the droplets. Integration then gives:
\begin{eqnarray}
H = 2a\cot\beta\left(\frac{\cosh(d/2a)}{\sinh(d/a)}\right).
\label{H}
\end{eqnarray}

\noindent For $d$ = 3 mm, $H \sim 300\, \mu$m, which is smaller, but on the order of magnitude of the drop size. When a particle deposited inside the channel reaches the edge, it slides down the liquid slope to reach the bath level. In the process, its potential energy is converted into kinetic energy so that $1/2 m (V_f^2 - V_i^2) = m g H$. For $H$ = 300 $\mu$m and $V_i$ = 3 cm/s, we expect $V_f$ to reach 6 cm/s, which close to our observations, see \ref{figure3}b. Using equation \ref{H}, we can thus predict the increase of the particles kinetic energy as they slide down the slope. In non-dimensional form, we find: 

\begin{eqnarray}
\frac{\Delta V^2}{2ga} = 2\cot\beta\left(\frac{\cosh(d/2a)}{\sinh(d/a)}\right)
\label{DeltaE_k}
\end{eqnarray}
 
\noindent Equation \ref{DeltaE_k} is reported in figure \ref{figure3}d as a dotted line, which is found to nicely match the experimental data, with a fitting parameter $\beta$ = 62$^{\circ}$, close to the value obtained on a single wall. 

%%%%%%%%%%%%%%%%%%% FIGURE 4 %%%%%%%%%%%%%%%%%%%%%%%%%%%%%
\begin{figure}[!t]% Use the figure* environment to get a wide figure \texttt{twocolumn} formatting. 
%\centering
\includegraphics[width=0.85\columnwidth]{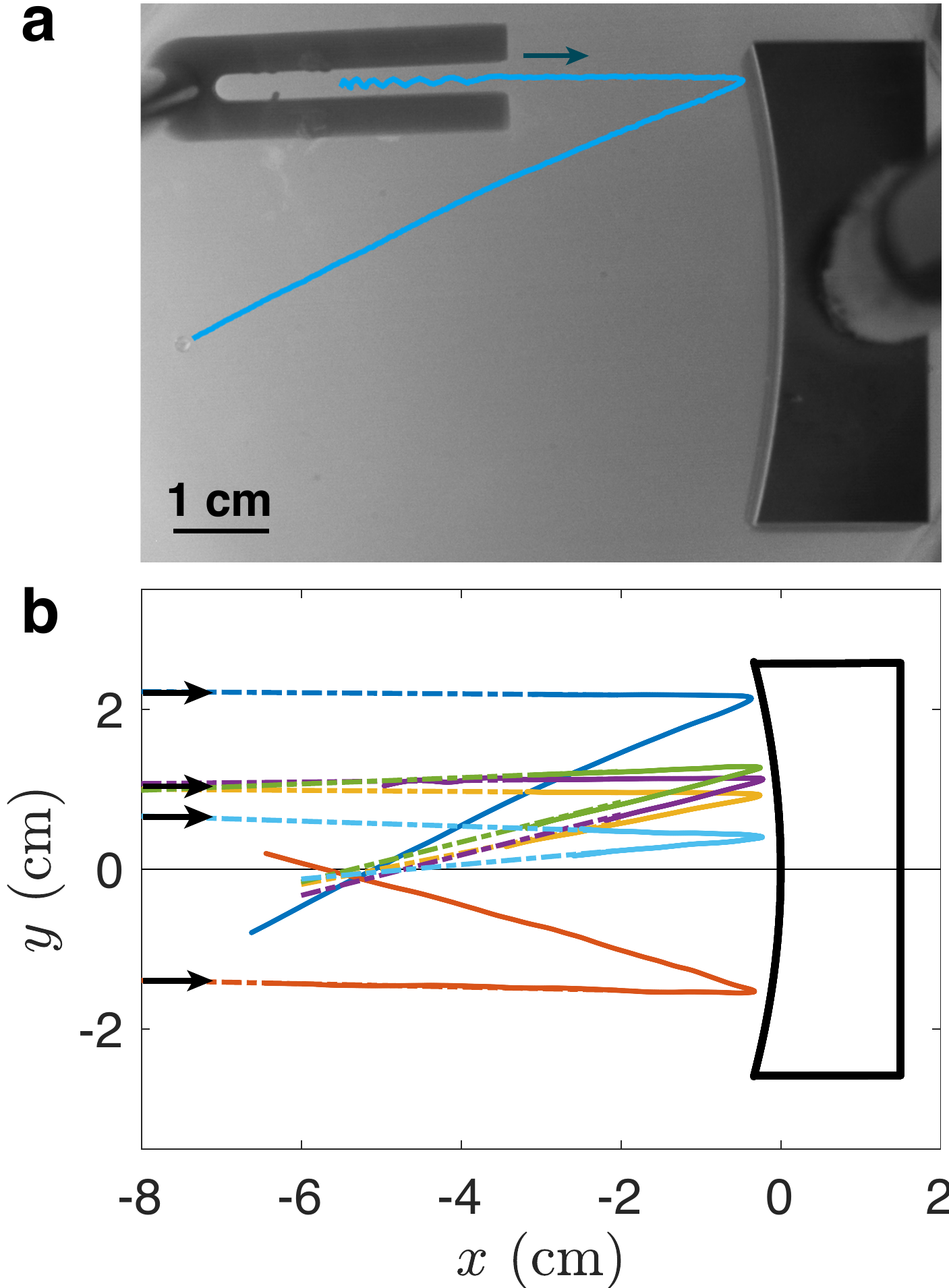}% Here is how to import EPS art
\caption{\label{figure4}\textbf{a.} Top view of the trajectory of a drop (radius $R$ = 830 $\mu$m) directed towards a parabolic mirror (with focal length 5 cm). See also Supplementary Movie 3. \textbf{b.} Trajectories of 6 drops, initially parallel to the mirror. They are all deflected towards the focal point.}
\end{figure}
%%%%%%%%%%%%%%%%%%%%%%%%%%%%%%%%%%%%%%%%%%%%%%%%%%%%%

\noindent
\textbf{Drop focusing.} These channels can be used in combination with other tools to finely control drop motion. We show an example of it in Figure \ref{figure4} and Supplementary Movie 3, where we use both a channel and a parabolic wall to focus drops in a specific location. This parabolic ``mirror" is visible on the right in figure \ref{figure4}a and it is designed with a focal length of 5 cm. Using a 3 mm wide channel, drops are launched parallel to the axis of the mirror and their trajectories are tracked before and after bouncing. The trajectory of a drop with radius $R$ = 830 $\mu$m is shown in blue in Figure \ref{figure4}a and a blue arrow indicates its initial direction of motion. The drop is reflected by the wall and exhibits a trajectory quite similar to what might be expected of light reflected on a parabolic mirror. To verify this, we repeated the same experiment multiple times by changing the position of the channel with respect to the focal axis of the parabola The corresponding drop trajectories are represented in Figure \ref{figure4}b. The continuous lines are the experimental trajectories (measured as soon as the drops leave the channel) and the dotted lines are their linear extrapolations before and after the rebound. On figure \ref{figure4}b, all drop trajectories cross the principal axis of the parabolic mirror at position $x =  5.4 \pm 0.3$ cm, a value that is very close to the actual focal length, efficiently gathering the droplets at this specific location. This experiment is similar in spirit to the deflection of gliding droplets by a cylinder, which was discussed recently by Hale and Boudreau \cite{Hale:2019}. By using a concave object instead of a convex one, droplets are scattered on the liquid surface instead of being focused, which is another nice demonstration of how capillary forces can be harnessed to direct drops.

\section{Conclusion} 
\noindent By studying the collision dynamics of a droplet with a single wall, we show that frictionless particles are a good tool to probe interfaces. The meniscus shape can be directly inferred from the variation of the particle's kinetic energy as a function of the distance to the wall. Here, we use liquid nitrogen, a cryogenic liquid with low surface tension and low latent heat of vaporization. Liquid nitrogen wets a large majority of surfaces, but the enhanced evaporation at the tip of the meniscus makes it particularly hard to determine its contact angle. Here, we obtain an effective contact angle $\theta$ = 30$^\circ$, surely dominated by the evaporation process more than by an equilibrium of surface tension forces. In addition, we show in the second part that, counter-intuitively, it is easier to control drops on a liquid bath than on solids. The menisci efficiently repel drops with extremely low energy loss (for our Leidenfrost droplets) and cause almost perfect reflections. In addition, drop ``waveguides" can be made by placing them between two walls. By tuning the distance between the walls, one can choose the amplitude of the acceleration that the drops undergo upon escape. To push this further, one could combine simple geometrical elements into more complex setups that could handle multiple drops at the same time and select, sort them and mix them.

\section*{Conflicts of interest}
The authors declare no conflict of interest. 

%\section*{Acknowledgements}
%The Acknowledgements come at the end of an article after Conflicts of interest and before the Notes and references.

%%%END OF MAIN TEXT%%%
\balance


\providecommand{\noopsort}[1]{}\providecommand{\singleletter}[1]{#1}%
\providecommand*{\mcitethebibliography}{\thebibliography}
\csname @ifundefined\endcsname{endmcitethebibliography}
{\let\endmcitethebibliography\endthebibliography}{}


%%%REFERENCES%%%
%\bibliography{biblio_menisci} %You need to replace "rsc" on this line with the name of your .bib file
\bibliographystyle{rsc} %the RSC's .bst file

\end{document}